# Tough self-healing elastomers by molecular enforced integration of covalent and reversible networks

Jinrong Wu [a, b], Li-Heng Cai [b, *], and David A. Weitz [b,*]

[a]State Key Laboratory of Polymer Materials Engineering, College of Polymer Science and Engineering, Sichuan University, Chengdu 610065, China.
[b]School of Engineering and Applied Sciences, Harvard University, Cambridge, MA 02138, USA.
*Correspondence and requests for materials should be addressed to: L.-H.C., lhcai@seas.harvard.edu or D.A.W., weitz@seas.harvard.edu

**Corresponding author contact:**
    Dr. David A. Weitz
    231 Pierce Hall
    School of Engineering and Applied Sciences
    Harvard University
    Cambridge, MA 02138
    Email: weitz@seas.harvard.edu
    Phone: 617-496-2842
    Fax: 617-495-0426

    Dr. Liheng Cai
    RM 517, Gordon McKay Laboratory
    School of Engineering and Applied Sciences
    Harvard University
    Cambridge, MA 02138
    Email: lhcai@seas.harvard.edu

**Word Count:**
    Number of text pages: 6
    Number of figures: 4
    Abstract: 191
    Main text: 2881
    Materials and Methods: 578
    Figure Legends: 723
    Total: 4373

**Abstract:** Self-healing polymers crosslinked by solely reversible bonds are intrinsically weaker than common covalently crosslinked networks. Introducing covalent crosslinks into a reversible network would improve mechanical strength. It is challenging, however, to apply this design concept to 'dry' elastomers, largely because reversible crosslinks such as hydrogen bonds are often polar motifs, whereas covalent crosslinks are non-polar motifs, and these two types of bonds are intrinsically immiscible without co-solvents. Here we design and fabricate a hybrid polymer network by crosslinking randomly branched polymers carrying motifs that can form both reversible hydrogen bonds and permanent covalent crosslinks. The randomly branched polymer links such two types of bonds and forces them to mix on the molecular level without co-solvents. This allows us to create a hybrid 'dry' elastomer that is very tough with a fracture energy 13,500J/m$^2$ comparable to that of natural rubber; moreover, the elastomer can self-heal at room temperature with a recovered tensile strength 4MPa similar to that of existing self-healing elastomers. The concept of forcing covalent and reversible bonds to mix at molecular scale to create a homogenous network is quite general and should enable development of tough, self-healing polymers of practical usage.

**Keywords:** tough; self-healing; supramolecular; molecular design; elastomers

*Main text*

Self-healing polymeric materials can revert to their original state with full or partial recovery of mechanical strength after damage.[1, 2, 3] As such, they hold great promise to extend the lifetime of polymeric products in many fields; examples include aerospace, automotive, civil, and medical engineering.[4] Unlike classical polymer networks that are crosslinked by permanent covalent bonds, self-healing polymeric materials are often based on reversible associations, such as hydrogen bonds, [3, 5] metal-ligand coordination,[6, 7] ionic interactions, [8] electrostatic interactions,[9] hydrophobic associations,[10] π–π stacking,[11] or polymer entanglements [2, 12]. Such reversible associations can break and reform to enable self-healing ability, but they are nevertheless relatively weak compared to covalent bonds. Thus, the toughness of self-healing polymers does not match that of covalent polymer networks such as natural rubber. Introducing permanent, covalent crosslinks into a reversible network would improve its mechanical properties, and this concept has been explored extensively to create tough hydrogels.[13, 14] [15] However, hydrogels contain a large amount of water that can evaporate, whereas diverse applications often require polymers that are not only tough, but also dry, such that they do not leach molecules and change properties. Unfortunately, it is challenging to integrate both covalent and reversible networks in a 'dry' polymer that does not contain co-solvents; reversible crosslinks such as hydrogen bonds are often polar motifs, whereas covalent crosslinks are non-polar motifs, and hence these two types of networks are intrinsically immiscible without co-solvents.[16] Consequently, despite its significant practical importance, it remains an unmet need for the development of elastomers with an exceptional combination of toughness and self-healing ability.

Here we report a tough, self-healing 'dry' elastomer that contains both covalent and reversible networks. The hybrid elastomer is fabricated by crosslinking a randomly branched polymer carrying motifs that can form both reversible hydrogen bonds and permanent covalent crosslinks. The randomly branched polymer links these two types of bonds and forces them to mix on the molecular level without macroscopic phase separation. This allows us to create a homogenous, optically transparent 'dry' elastomer without co-solvents. At small deformations, the hydrogen bonds break and reform to dissipate energy. At large deformations, the hybrid elastomer exhibits patterns that are reminiscent of crazes observed in typical plastics, but are of much larger length scales, ranging from 1 to 1000 μm. These patterns are so-called 'macro-crazes'; they are unique to the hybrid elastomer and help maintain material integrity at large deformations. The ability to sacrifice hydrogen bonds at small deformations and maintain material integrity at large deformations enables a very tough elastomer with fracture energy 13,500J/m$^2$ comparable to that of natural rubber. Moreover, the hybrid elastomer self-heals at room temperature with a recovered tensile strength of 4 MPa, which is comparable, if not better, to the existing self-healing elastomers.

We first synthesize a randomly branched polymer carrying motifs that can form both reversible hydrogen and permanent covalent bonds. To this end, we use diamine molecules and acrylic acid as raw reagents, and keep the molar ratio between amine and carboxyl groups fixed at 1:1.25 (**Materials and Methods**). A diamine molecule has an amine group at each of its two ends. An amine group reacts with the unsaturated C=C in acrylic acid at 50 ℃ through Michael addition, resulting in carboxyl functionalized molecules, as described by Step 1 in **Figure 1a** and detailed in **Figure S1**. A carboxyl group can further react with an amine group to form an amide group at relatively high temperature, 150 ℃, through condensation reaction, as described by Step 2 in **Figure 1a**. We allow the condensation reaction to proceed for about 4 hours to form randomly branched polymers. In a randomly branched polymer, the branching point is a tri-functional linker, the end groups are either carboxyl or primary amine groups, and along the backbone of branches are randomly distributed amide groups, as illustrated in **Figure 1b**. The chemical reactions and resulted functional groups are confirmed by Fourier transformation infrared spectroscopy (FTIR) and nuclear magnetic resonance (NMR) measurements (**SI Materials and Methods** and **Figures S2&S3**).

The randomly branched polymers form a supramolecular network that is connected by either amide-amide or carboxyl-amine hydrogen bonds. Unlike permanent covalent bonds, these hydrogen bonds are temporary and can dissociate at room temperature; consequently, the network is a viscous liquid rather than a solid, as evidenced by the dynamic mechanical test in which the loss modulus is always higher than the storage modulus in a wide range of shear frequency (**Figure S4**). Moreover, the liquid becomes much less viscous at higher temperature, reflected by a dramatic decrease in loss modulus, as shown by the open symbols in **Figure S4**. This liquid-like feature is critical for the fabrication of a hybrid network. It promotes the mobility of reactive amine and carboxyl groups, and therefore, facilitates the condensation reaction between them, through which the precursor polymers are crosslinked to form a network. Moreover, the liquid-like feature allows the network to be molded to desired shapes.

After obtaining the viscous liquid, we transfer it into a Teflon mold, and maintain the temperature at 160 ℃ for 32 hours. During this process, amine groups and carboxyl groups at the ends of branching arms react with each other through condensation reaction. This crosslinks randomly branched polymers to form a network, as described by Step 3 in **Figure 1a**. The network contains both reversible hydrogen bonds and permanent crosslinks, as illustrated by **Figure 1c**; thus, we call it a hybrid network. The network is an optically transparent solid, as shown by an optical image in the inset of **Figure 1d**. Such optical transparency reflects homogeneity of the network at molecular length scale; otherwise the intrinsic immiscibility between polar hydrogen bonds and non-polar covalent crosslinks would result in visible phase separation and optical turbidity. The homogeneity of the network is further demonstrated by small angle X-Ray scattering, which does not exhibit characteristic peaks associated with micro-phase separation (**Figure S5**). Moreover, the polymer network has a glass transition temperature of 4-14 ℃ that is below room temperature; therefore, the material is an elastomer rather than a plastic at room temperature (**Figure 1d**).

To characterize mechanical properties of the hybrid elastomer, we cut the material into dumbbell shape and quantify its stress-strain behavior using uniaxial tensile test (**Materials and Methods**). In a typical test, we stretch the sample at a controlled strain rate of 0.014/sec, and simultaneously monitor the elongation of the sample using a camera. Example images of the hybrid elastomer under different extents of strain are shown in **Figure 2a**, and a representative movie for a tensile test is shown in **Movie S1**. Examining the dependence of stress on strain reveals three distinct regimes for mechanical response. *Regime 1:* At small strain, $\epsilon < 0.04$, the stress increases rapidly with strain in a linear manner (light green area and left inset in **Figure 2b**). *Regime II:* At intermediate strain, the rate of increase in stress slows down, yet maintains a nearly linear behavior (light red area in **Figure 2b**). *Regime III:* At large strain, $\epsilon > 0.98$, the elastomer exhibits strain stiffening with the rate of increase in stress becomes higher compared to that in Regime II (light blue area in **Figure 2b**). Interestingly, the onset of Regime III is featured by an abrupt decrease in stress, as shown by the right inset in **Figure 2b**. Moreover, this decrease in stress is accompanied by the appearance of white lines that are perpendicular to the direction of elongation, as shown by the optical image of the sample at $\epsilon = 0.98$ in **Figure 2a**. These white lines are reminiscent of crazing in plastics, which is due to localized large strain around defects[17], but are different in that they are distributed over the whole sample rather than being localized. Even more striking is the length scale on which they occur; the crazes are remarkably large and visible by eye, in contrast to conventional, micro-scale crazes in plastics. We therefore name these white lines 'macro-crazes'.

To explore the microstructure of these 'macro-crazes', we examine them *in situ* using scanning electron microscopy (SEM) (**Materials and Methods**). To do so, we introduce a notch on a rectangular sample, and monitor the formation of 'macro-crazes' near the notch as the sample is subjected to extension. Initially, the hybrid elastomer exhibits no 'macro-crazes' except the crack due to initial notch, as shown in **Figure 2c(i)**. Upon stretching, 'macro-crazes' are initiated and extended from the notch zone across the sample until rupture (**Figure 2c(ii)**). The 'macro-crazes' are small gaps with a wide distribution of widths ranging from ~1μm to ~1000μm, each connected by stretched lamellae aligned along the direction of elongation, as shown by the SEM images in **Figure 2c(iii-iv)**. By contrast, conventional elastomers crosslinked solely by covalent bonds or supramolecular networks crosslinked solely by reversible bonds do not exhibit crazing upon deformation. Moreover, the hybrid elastomer maintains its integrity in the presence of a large amount of 'macro-crazes' at relatively large deformation, as shown by the images at $\epsilon > 0.98$ in **Figure 2a**. The ability of the hybrid elastomer to withstand large deformation without breaking is likely due to the reversible bonds, which can break and reform, redistributing stress across the whole material and delaying accumulation of local stress.

Unlike covalent bonds, reversible hydrogen bonds have a finite lifetime and can relax. Therefore, the mechanical properties of the hybrid elastomer depend on the rate of deformation. To explore this, we perform tensile tests at different strain rates for parallel hybrid elastomer samples. We vary the strain rate by nearly three orders of magnitude, from 0.0014/sec to 0.68/sec. For all strain rates explored, the stress-strain behavior is characterized by three regimes depending on the extent of deformation, as shown in **Figure 3a**. In each regime, the stress increases nearly linearly with strain, and the rate of increase describes the Young's modulus of the material. Interestingly, at small deformation in Regime I the Young's modulus is strain rate dependent. At small strain rates, $\dot{\epsilon} < 0.01/sec$, the Young's modulus in Regime I, $E_I$, increases rapidly with strain rate, following a power law, $E_I \propto \dot{\epsilon}^{0.65}$, from approximately 10MPa to 100MPa; at larger strain rates, $\dot{\epsilon} > 0.01/sec$, $E_I$ saturates at approximately 100MPa, as shown by the circles in **Figure 3b**. The crossover strain rate is about 0.03/sec. Qualitatively such behavior can be understood based on the relaxation of reversible hydrogen bonds. A hydrogen bond has a finite average lifetime, and the lifetimes of all hydrogen bonds are randomly distributed around this mean.[18] Thus, a reversible bond can break at any time during the experiment, but is less likely at time scales shorter than the average lifetime. At shorter time scales, or at higher strain rates, the fraction of un-relaxed reversible bonds is larger than that at longer time scales. Consequently, the contribution to the network modulus due to reversible bonds increases with strain rate. At very high strain rates, however, this increase becomes saturated because nearly all hydrogen bonds are un-relaxed; therefore, the network moduli in the high strain rate limit become nearly a constant. Indeed, the crossover strain rate, 0.03/sec, is on the same order of magnitude as the reciprocal of the average lifetime of an amide-amide hydrogen bond[19].

Such understanding of strain rate dependent network modulus at small deformation is further supported by dynamic mechanical measurements (**Materials and Methods**). In a typical measurement, we quantify the network shear modulus at small strain, 0.05%, and oscillatory shear frequency from 0.01Hz to 100Hz. We perform such measurements at different temperatures, and construct a master curve using classical time-temperature superposition[20]; this allows us to probe the network shear modulus over an extremely wide range of frequencies. We find that the shear storage modulus, *G'*, increases weakly with frequency, $\omega$, following a power-law, $G'(\omega) \propto \omega^{0.07}$. As $\omega$ increases from $10^{-11}$Hz to ~1Hz, the storage modulus increases by nearly one order of magnitude from ~3MPa to ~30MPa, as shown by the red symbols in **Figure 3c**. Since the shear modulus is 1/3 of the Young's modulus for an elastomer with a Poisson ratio of 0.5, such increase in shear modulus is in quantitative agreement with that observed in uniaxial tensile tests, in which the Young's modulus at small deformation increases from 10MPa to 100MPa as the strain rate increases (the circles in **Figure 3b**). Moreover, at the lowest frequency, the magnitude of shear modulus, 3MPa, is consistent with equilibrium relaxation modulus, 2.5MPa, as shown by the stress relaxation curve in **Figure S6**. Together these measurements imply that the relaxation of reversible bonds determines the rate-dependent mechanical properties of the hybrid network at very small deformation.

At relatively large deformations, the mechanical properties of the hybrid network have a weak dependence on the strain rate: For intermediate deformations in Regime II, the Young's modulus increases with strain rate following a very weak power law, $E_{II} \propto \dot{\epsilon}^{0.18}$ (squares in **Figure 3b**); for large deformations in Regime III, the Young's modulus is essentially independent of strain rate, $E_{III} \propto \dot{\epsilon}^{0.04}$ (diamonds in **Figure 3b**). The weak dependence in each regime is likely the result of most hydrogen bonds being broken under such large deformations; thus the major contribution to the network modulus is from permanent bonds, which do not change with strain rate. Interestingly, the onset of Regime III, which is associated the initiation of 'macro-crazes', is nearly independent of strain rate, as shown in **Figure 3d**. Upon further deformation, these 'macro-crazes' continue to grow and help redistribute stress to maintain material integrity.

The ability to sacrifice hydrogen bonds at small deformation and maintain material integrity at large deformation allows the hybrid network to dissipate energy at multiple length scales. To quantify the efficiency of energy dissipation of these hybrid elastomers, we perform cyclic tensile tests at a fixed strain rate of 0.014/sec and at different deformations. In a typical test, there is a large hysteresis during the loading and un-loading processes, as shown by the stress-strain curve in **Figure 4a(i)**. We define the efficiency of energy dissipation as the ratio between the integrated area in the hysteresis loop and that under the loading curve, and find that the efficiency is remarkably high, with a value of ~75%; moreover, such high efficiency is nearly the same regardless of the extent of deformation, up to strains of 1.4, as shown in **Figure S7**. This suggests that the hybrid network is very dissipative, a characteristic feature of typical tough materials.[14] To quantify the fracture toughness of the hybrid network, we perform tensile tests on single-edge notched samples at a strain rate of 0.005/sec at room temperature, and calculate the fracture energy $G_c$ using the Greensmith method (see **SI Materials and Methods**)[21]. For the hybrid network, $G_c$ can reach values as high as 13.5kJ/m$^2$ (filled circles in **Figure 4c**), which is comparable to that of natural rubber under similar testing conditions.[22] Moreover, the toughness of the hybrid network is more than twice of recently reported tough interpenetrating dry networks (empty squares in **Figure 4c**) [23].

Unlike interpenetrating elastomers formed by solely covalently crosslinked networks[23], these hybrid elastomers can self-heal because of reversible hydrogen bonds. For example, under intermediate deformations the hybrid elastomer recovers to its original mechanical properties after waiting for 10 mins or longer, as shown by stress-strain curves in **Figure 4a(ii)**. Moreover, even after being cut into two pieces, the hybrid network can partially recover its mechanical strength. After bringing two freshly fractured surfaces into contact at room temperature, and waiting for 1 hour, the sample is able to support a stress of 2MPa. After 12 hours, the sample recovers a tensile strength of 4MPa, about 30% of its original value. Fracturing the sample breaks both the covalent and reversible bonds; only the reversible are reformed, resulting in the reduced value of the recovered tensile strength. Thus, the degree of healing of this hybrid elastomer is lower than that of self-healing 'dry' polymers which are formed solely with reversible bonds. Nevertheless, the absolute magnitude of the recovered tensile strength, 4MPa, of the hybrid elastomer is comparable to the pristine tensile strength of most self-healing elastomers.[3, 7, 24]

In summary, we have developed a hybrid elastomer by crosslinking randomly branched polymers carrying motifs that can form both reversible hydrogen bonds and permanent covalent crosslinks. The randomly branched polymer links such two types of bonds and forces them to mix on the molecular level, producing a homogenous, optically transparent elastomer. At small deformations, the reversible bonds break and reform to dissipate energy; at relatively large deformations, covalent bonds start to break, as illustrated in **Figure 4d**. However, the material maintains its integrity at large deformations because reversible bonds help redistribute stress; this feature is associated with 'macro-crazing' that is unique to this hybrid elastomer. Such synergy of energy dissipation at both molecular and macroscopic length scales produces an extremely tough elastomer that has fracture toughness comparable to natural rubber. However, the molecular mechanisms of 'macro-crazing' at large deformations, and how this behavior contributes to energy dissipation of the network, remain to be explained. Exploring these questions will require model networks of more controlled network structure and molar ratio between reversible and covalent crosslinks. Nevertheless, the combination of optical transparency, toughness, and self-healing ability will enable applications of the hybrid elastomers in stretchable electronics and damping materials.[25] Moreover, the concept of using molecular design to mix covalent and reversible bonds to create a homogenous hybrid elastomer is quite general and should enable development of tough, self-healing polymers of practical usage.

**Materials and Methods**

**Reagents.** All chemicals were used as received unless otherwise noted. Acrylic acid (AA, 99%, ~ stabilized with 200ppm 4-methoxyphenol), 1, 12-diaminododecane (DADD, 99+%) and chloroform (99.8+%, stabilized with ethanol) were purchased from Alfa Aesar; 1, 6-hexamethylenediamine (HMDA, 98%) and N, N-dimethyl formamide (DMF, 99.8%) were purchased from Sigma Aldrich.

**Synthesis.** We synthesized the hybrid networks by two steps, pre-polymerization and crosslinking. During the pre-polymerization, 0.125 mole AA was added dropwise into 100 mL chloroform solution of DADD and HMDA under agitation at 50 °C in a three-neck round bottom flask. The molar ratio of -COOH in AA to $-NH_2$ in diamine was fixed at 1.25:1, whereas that of HMDA/DADD was varied between 0.3/0.7 and 0.5/0.5. After the addition of AA, we stirred the solution at 50 °C for 16 hours, and then heated it up to 80 °C and maintained the temperature for about 3 hours under nitrogen flow to remove chloroform. Subsequently, we raised the temperature to 150 °C and maintained it for about 4 hours. The temperature was then lowered to 140 °C, and 50 ml DMF was charged into the flask. This results in a solution that contains randomly branched polymers. During the crosslinking step, we casted an aliquot of the solution in a Teflon mold, and placed this mold in a glass reactor with nitrogen inlet and outlet. We heated the glass reactor to 110 °C and maintained the temperature overnight to slowly evaporate DMF. The temperature was then increased to 160 °C with a rate of 10 °C per hour, and was maintained at 160 °C for 32 hours, resulting in formation of a covalently crosslinked elastomer sheet with a thickness of ~1mm.

**Bulk rheology.** Bulk rheological measurements were carried out on a strain controlled rheometer (ARES-G2, Texas Instruments) with 5mm plate-plate geometry. We cut a hybrid elastomer sheet into a circular shape with diameter of 5mm using a punch. The circular sample was glued to the plates to avoid slippage. Frequency sweeps were performed from $10^2$ to $10^{-2}$ Hz at 0.5% stain at temperatures of 20 °C, 40 °C, 60 °C, 80 °C, 100 °C, 120 °C, and 130 °C.

**Mechanical test.** We characterized the mechanical properties of hybrid elastomers using Instron® 3342 with a 1000N load cell. Uniaxial tensile measurements were conducted at room temperature in air under different strain rates, including 0.0014, 0.014, 0.14, 0.41 and 0.68 /sec. We measured the strain by monitoring the displacement of two markers in the central part of a dumbbell shaped sample using a camera. Each measurement was repeated at least three times. Cyclic extension tests with incremental elongation were performed on the same tensile machine at a strain rate of 0.014/sec. The maximum strains of the cyclic tensile tests include 0.3, 0.5, 0.8, 1.1 and 1.4.

**Scanning electron microscopy.** We monitored the growth of 'macro-crazes' by performing tensile test in the chamber of an environmental scanning electron microscopy (SEM, Zeiss EVO 55). A notch of about 0.4 mm in length was made on a specimen of 4 mm width and 1mm thickness. The specimen was coated with a Pt/Pd layer of 5nm thickness and then fixed on a screw-driven tensile stage. The tensile stage was then placed inside the chamber of SEM, and images were taken around the tip of the notch. The stress and strain of the sample were continuously recorded at a strain rate of about 0.014/sec. Using SEM at an acceleration voltage of 5 kV, we monitored the formation of the 'macro-crazes' during the tensile process until the sample was broken.

**Supporting Information**

Supporting Information is available from the Wiley Online Library or from the author.

**Acknowledgement**

This work is supported by NSF DMR-1310266, the Harvard MRSEC DMR-1420570, and NIH/NHLBI 5P01HL120839-03, and in part by Capsum under contract A28393. J.W. is supported by the NSFC 51673120. We thank Prof. Alfred Crosby, Prof. Zhigang Suo, Prof. Chinedum Osuji, and Prof. Cornelis Storm for enlightening discussions.

**Figures**

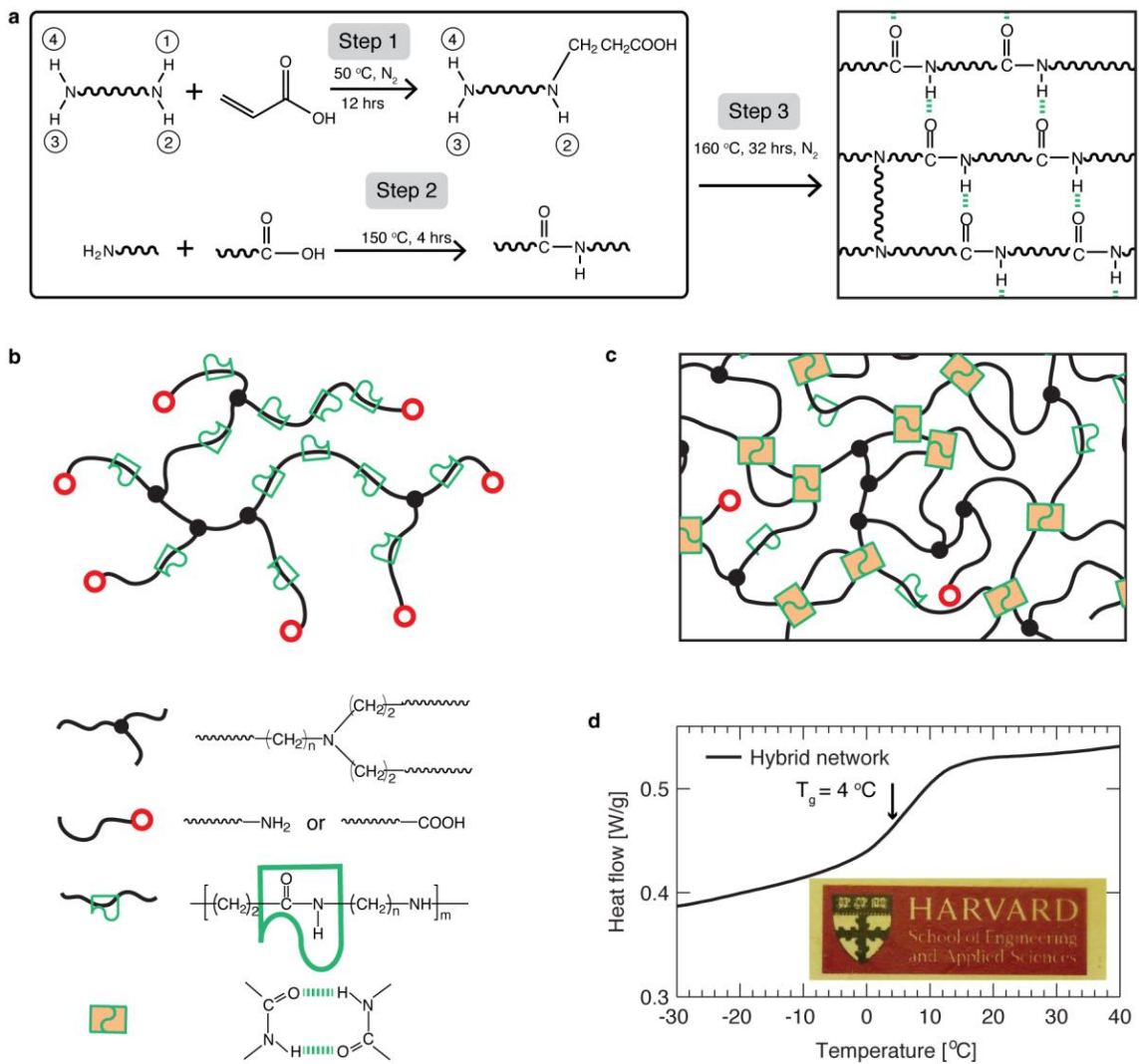

**Figure 1. Concept and synthesis of hybrid networks. a,** Synthesis of a hybrid network. Step 1: The double bond of an acrylic acid molecule reacts with one of the four N-H bonds in a diamine molecule through Michael addition, resulting in an oligomer with one end functionalized by a carboxyl group, whereas the rest three N-H bonds are still reactive to double bonds. There are four reactive hydrogen groups in a diamine molecule, and depending on the number of reacted hydrogen groups, five types of oligomers can be generated (**Figure S1**). Step 2: Polymerization of the oligomers formed in Step 1 through condensation reaction between the amine groups and carboxyl groups at 150 °C; this forms a randomly branched polymer. Step 3: The randomly branched polymers are crosslinked to form a network through condensation between the amine groups and carboxyl groups. **b,** Illustration of a randomly branched polymer formed after Step 2. The polymer carries motifs that can form reversible bonds (green symbols) along the backbone of polymers and motifs that can form covalent bonds (red circles) at the end

of branching arms. Filled black circles represent tri-functional covalent branching points. **c,** Illustration of the hybrid network formed after Step 3. The formation of covalent network is achieved by linking the end groups of the randomly branched polymer by condensation reaction, and the formation of reversible associations is illustrated by the change of a pair of empty green symbols to filled ones. The empty red circles represent carborxyl or amine groups, and the green empty symbols represent an amide group. **d,** Differential scanning calorimetry (DSC) analysis of a hybrid network. The glass transition temperature of the network is 4$^{o}$C, indicating the material is an elastomer rather than plastic at room temperature. Inset: a representative image of a hybrid network.

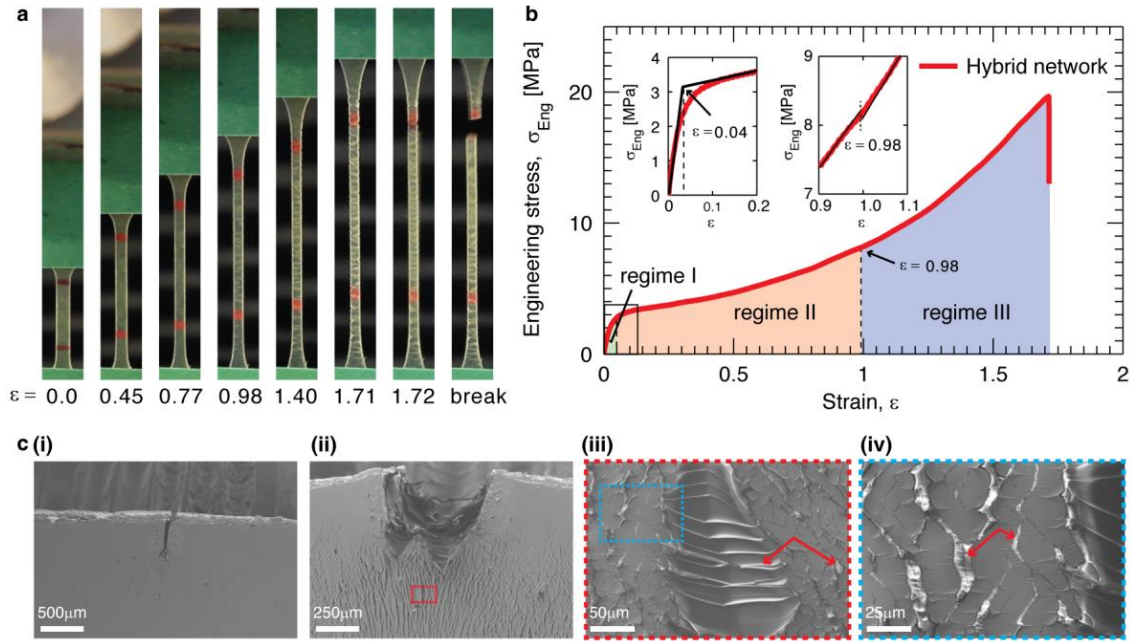

**Figure 2. Strain dependent mechanical properties. a,** Optical images of a hybrid elastomer at different strains, ε, when subjected to uniaxial tensile test. The hybrid elastomer is fabricated using a hexamethylenediamine/diaminododecane molar ratio 0.5/0.5 (**Materials and Methods**). **b,** Dependence of engineering stress, $\sigma_{eng}$, on the strain. The stress-strain curve can be divided into three regimes depending on the extent of strain: Regime I, ε<0.04; Regime II, 0.04<ε<1.0; Regime III, ε>1.0. The initiation of Regime III is associated with two features: (i) Mechanically the stress exhibits a sudden decrease (right inset), and (ii) visually the appearance of white imprints that are perpendicular to the direction of elongation at strain of 0.98 (**Fig. 2a**). **c,** Scanning electron microscopy (SEM) images of the hybrid network with an initial notch under uniaxial tensile test *in situ*.

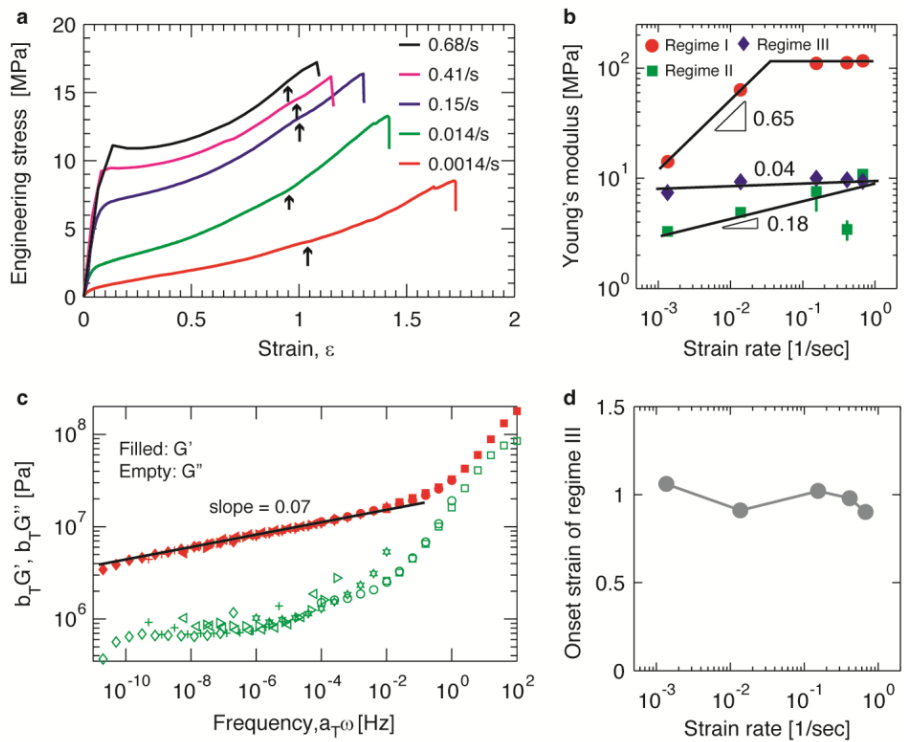

**Figure 3. Mechanical properties of hybrid elastomers at different strain rates. a,** Stress-strain curves at different strain rates for the hybrid elastomer with a hexamethylenediamine/diaminododecane molar ratio of 0.3/0.7. Arrows correspond to the strain at which the white lines appear. **b,** Young's modulus of hybrid elastomers at different strain rates in three regimes: small deformation (*Regime I*, circles), intermediate deformation (*Regime II*, squares), and large deformation (*Regime III*, diamonds). **c,** Frequency dependence of the storage (G', filled symbols) and loss (G'', empty symbols) moduli of hybrid elastomer obtained by classical time-temperature superposition. The reference temperature is 20 ℃ (squares), and measurements were performed at 40 ℃ (circles), 60 ℃ (hexagons), 80 ℃ (right triangles), 100 ℃ (left triangles), 120 ℃ (plus symbols), and 130 ℃ (diamonds). **d,** Dependence of onset strain for Regime III on strain rate.

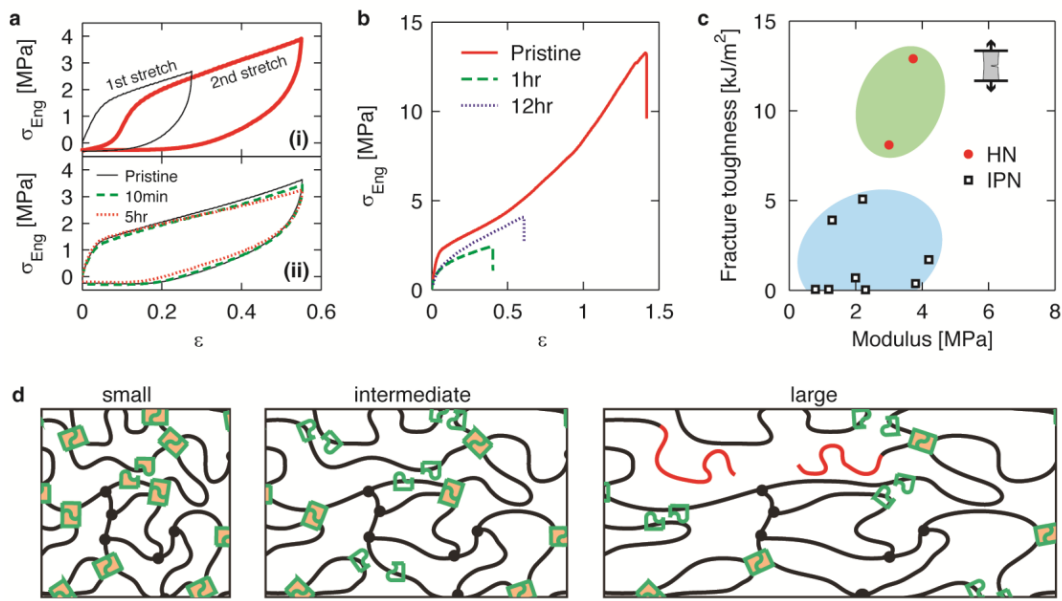

**Figure 4. Self-healing properties and toughness of hybrid networks. a,** Hybrid network completely heals in Regime II under intermediate strain. **(i)** Thin, black line represents the first cyclic stretch, and the thick, red line represents the second cyclic stretch immediately after the first one. **(ii)**, Solid line corresponds to the first stretch with strain up to that in Regime II; dashed line represents the second load after 10mins for the first load; dotted line corresponds to the cyclic test after waiting for 5 hours. **b,** Self-healing of hybrid networks in *Regime III*. Stress-strain curves for a pristine sample (red solid line), and parallel samples that are brought into contact for 1 hour (green dashed line) and for 12 hours (blue dotted line) after being cut into two parts. **c,** Comparison of toughness for hybrid networks (filled circles) and a recently developed tough interpenetrating networks (IPN) (open squares) [23]. **d,** Illustration of molecular structures for the hybrid elastomer under different extents of deformation.

# Table of contents

**A tough, self-healing polymer network** is fabricated by crosslinking randomly branched polymers carrying motifs that can form both reversible hydrogen bonds and permanent covalent crosslinks. The randomly branched polymer forces such two types of bonds to mix on the molecular level without co-solvents; this enables a hybrid 'dry' elastomer that with an exceptional combination of toughness and self-healing ability.

**Keywords:** tough; self-healing; supramolecular; molecular design; elastomers

**Authors:** Jinrong Wu [a, b], Li-Heng Cai [b, *], and David A. Weitz[b,*]

**Title**: Tough self-healing elastomers by molecular enforced integration of covalent and reversible networks

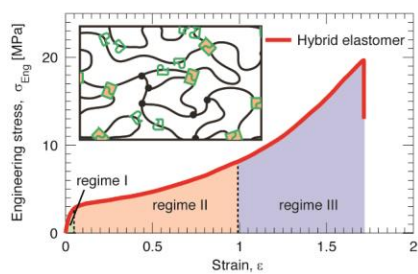

*Supporting Information*

# Tough self-healing elastomers by molecular enforced integration of covalent and reversible networks

Jinrong Wu [a, b], Li-Heng Cai [b, *], and David A. Weitz[b,*]

[a]State Key Laboratory of Polymer Materials Engineering, College of Polymer Science and Engineering, Sichuan University, Chengdu 610065, China.

[b]School of Engineering and Applied Sciences, Harvard University, Cambridge, MA 02138, USA.

*Correspondence and requests for materials should be addressed to: L.-H.C., lhcai@seas.harvard.edu or D.A.W., weitz@seas.harvard.edu

**Supporting Information**
SI Materials and Methods
Figures S1-S7
SI Movies

**SI Materials and Methods**

**S1. Materials.** Acrylic acid (AA, 99%, stabilized with 200ppm 4-methoxyphenol), 1, 12-diaminododecane (DADD, 99+%) and chloroform (99.8+%, stabilized with ethanol) were purchased from Alfa Aesar; 1, 6-hexamethylenediamine (HMDA, 98%) and N, N-dimethylformamide (DMF, 99.8%) were purchase from Sigma Aldrich. All the chemicals were used as received without further purification.

**S2. Synthesis of hybrid elastomers.** The elastomers were synthesized by two steps, including pre-polymerization and crosslinking. During the pre-polymerization, 0.125 mole AA was added dropwise into 100 mL chloroform solution of DADD and HMDA under agitation at 50 ℃ in a three-neck flask. The molar ratio of -COOH in AA to –$NH_2$ in diamine was fixed at 1.25:1, and the molar ratio of HMDA/DADD was varied between 0.3/0.7 and 0.5/0.5. After the addition of AA, the solution was stirred at 50 ℃ for 16 hours. During this process the vinyl groups of AA react with primary amine groups of diamines through Michael addition, forming secondary amine groups; these secondary amine groups are still reactive and can react with vinyl groups to form tertiary amine groups. After the completion of Michael addition, all the vinyl groups of AA are consumed, producing a mixture that contains five types of amidoamine oligomers carrying both carboxyl and amine groups, as shown in Step 1 in **Figure S1**. The mixture was then heated up to 80 ℃ and maintained for about 3 hours under nitrogen flow to remove chloroform. Subsequently, the temperature was raised up to 150 ℃ to enable the condensation reaction between carboxyl groups and amine groups of the oligomers and unreacted diamines, as shown in Step 2 in **Figure S1**. As the reaction proceeds to a certain extent (about 4 hours), the mixture becomes viscoelastic yet without solidifying. Before the mixture starts to rise up the stirring stem, the temperature was lowered to 140 ℃. 50 ml DMF was charged into the flask to dissolve the pre-polymer. The pre-polymer is a hyper-branched polymer with tertiary amines serving as branching points and reactive end groups of carboxyl groups and primary amine groups, as show in **Figure S1**.

During the crosslinking step, an aliquot of the pre-polymer solution was casted in a Teflon mold. This mold was placed in a glass reactor with nitrogen inlet and outlet. The glass reactor was heated at 110 ℃ for overnight to slowly evaporate DMF. Then the temperature was raised by 10 ℃ increment every hour to 160 ℃ to allow condensation reaction between carboxyl groups and amine groups of the hyper-branched polymer. The reaction was maintained at 160 ℃ for 32 hours to form a covalently crosslinked elastomer sheet with a thickness around 1mm.

**S3. FTIR characterization.** To confirm the Michael addition between double bonds and amine groups and the condensation reaction between carboxyl and amine groups, Fourier transform infrared spectroscopy (FTIR) spectra were collected on a Perkin Elmer FTIR spectrometer fitted with an attenuated total reflectance cell. The FTIR spectra of AA, HMDA, DADD and hybrid network elastomer are shown in **Figure S2**. AA exhibits characteristic peaks at 1700, 1635, 1298 and 1241 cm$^{-1}$, corresponding to C=O stretching, C=C stretching, C-O stretching coupled with O-H in plane bending, and O-H in plane bending coupled with C-O stretching, respectively. HMDA and DADD have characteristic peaks at 3329, 3256, 3166 and 1607 cm$^{-1}$, corresponding to N-H stretching of carbamate formed by amine capturing $CO_2$, asymmetric N-H stretching, symmetric N-H stretching, and N-H in plane bending, respectively. These characteristic peaks of both reactants disappear in the hybrid network, suggesting that the Michael addition and condensation reaction nearly consume all the vinyl groups, carboxyl groups and primary amine groups. The condensation between carboxyl groups and amine groups forms amide groups, as confirmed by the appearance of the characteristic amide I band at 1650 cm$^{-1}$ and amide II band at 1553 cm$^{-1}$ of the hybrid network.

**S4. NMR characterization.** To investigate the molecular structure of the hyper-branched pre-polymer, its $^1$H and $^{13}$C nuclear magnetic resonance (NMR) spectra were obtained using a Bruke Avance 400 spectrometer, as shown in **Figure S3**. The pre-polymer was dissolved in $D_2O$ and tested at room temperature. The NMR results further confirm the Michael addition and condensation reaction, and the formation of the hyper-branched pre-polymer. In particular, the branching point is a tertiary amine formed by reacting two double bonds with a primary amine or by reacting a carboxyl group with a secondary amine; while the terminal groups consist of primary amines and carboxyl groups.

$^1$H NMR (400 MHz, $D_2O$) δ (ppm) 1.25-1.62 (m, -CH$_2$-C*H*$_2$-CH$_2$-), 2.37-2.40 (t, -CH$_2$-C*H*$_2$-COOH, -CH$_2$-C*H*$_2$-N(CH$_2$-)-CH$_2$-), 2.49-2.51 (m, -CH$_2$-C*H*$_2$-NH-CH$_2$-, -NH-CO-C*H*$_2$-CH$_2$-N(CH$_2$-)-CO-,), 2.71 (m, -CO-CH$_2$-C*H*$_2$-N(CH$_2$-)-CH$_2$-, -CH$_2$-C*H*$_2$-NH$_2$), 2.83 (m, -NH-C*H*$_2$- CH$_2$-COOH), 2.92-.296 (m, -NH-C*H*$_2$-CH$_2$-CO-NH-), 3.12 (m, -CH$_2$-C*H*$_2$-NH-CO-CH$_2$-), 3.73-3.78 (t,-CH$_2$-C*H*$_2$-N(CO-)-CH$_2$-).

$^{13}$C NMR (400 MHz, $D_2O$) δ (ppm) 176.65-180.48 (-*C*O-NH-, -*C*OOH), 58.84 (-CH$_2$-*C*H$_2$-N(CH$_2$-CH$_2$-CO-)-CH$_2$-CH$_2$-CO-), 47.73 (-CH$_2$-N(*C*H$_2$-CH$_2$-CO-)-*C*H$_2$-CH$_2$-CO-, -CH$_2$-*C*H$_2$-N(CO-CH2-)-CH2-), 44.31 (-CH$_2$-*C*H$_2$-NH-CO-, -CH$_2$-NH-*C*H$_2$-CH$_2$-CO-, -*C*H$_2$-NH$_2$), 38.55-39.94 (-CH$_2$-*C*H$_2$-CO-NH-, -CH$_2$-*C*H$_2$-COOH), 32.94 (-*C*H$_2$-CH$_2$-NH$_2$, -*C*H$_2$-CH$_2$-NH-), 25.02-28.63 (-CH$_2$-*C*H$_2$-CH$_2$-).

**S5. Rheological measurements.** *Melt of randomly branched polymers.* Rheological experiments were carried out on a stress controlled rheometer (MCR501, Anton Paar) with 25 mm plate–plate geometry at a gap of 750 μm. Frequency sweeps were performed from $10^2$ to $10^{-2}$ Hz at 0.5% stain at temperatures ranging from 40 ℃ to 100 ℃. Changes in normal force due to a gap contraction with temperature were alleviated by adjusting the gap height.

*Hybrid elastomers.* To measure the equilibrium modulus of the hybrid elastomer, we quantify its stress relaxation at elevated temperature, 100 ℃; this allows the hybrid elastomer to relax relatively fast such that it can reach equilibrium relaxation during experimental time scale. In particular, we laser cut a hybrid elastomer sheet into a disk with diameter of 25mm, and mount the sample between a plate-plate geometry using a thin layer of glue. After the glue is solidified, we raise the temperature to 100 ℃. Then we apply a step strain of 1.0%, and monitor how the stress decays with time, as shown by the dots in **Figure S6**. We fit the data using an empirical equation, $G(t) = G_0 \left[1 + \left(\frac{t}{\tau}\right)^{-\alpha}\right]$, in which $G_0$ is the network equilibrium modulus, $\tau$ is the characteristic relaxation time of the network, and $\alpha$ is the exponent that describes how fast the network relaxes. The best fit is shown by the line in **Figure S6**, which gives the equilibrium shear modulus of 2.35 MPa.

**S6. SAXS measurement.** To test the phase morphology of the hybrid network elastomer, small angel X-ray scattering (SAXS) measurement was performed on XEUSS 2.0 with a wavelength of 0.154nm under vacuum. Two-dimensional SAXS pattern was acquired using a Pilatus SAXS detector. The exposure time was 300s. A silver behenate (AgC22H43O2) standard was used to calibrate the scattering angle. The sample-to-detector distance was 2479 mm. The diffraction and scattering signals were corrected for beam fluctuation and background scattering. X-ray data analysis was performed using the Fit2D software to acquire SAXS intensity vs. scattering vector curve, as shown in **Figure S4.** The absence of scattering maximums suggests that there is no phase separation in the hybrid network elastomer.

**S7. Differential scanning calorimetry (DSC) measurements.** To determine the glass transition temperature of the hybrid network elastomers, differential scanning calorimetry (DSC) was performed using Q200 (TA Instruments). The mass of the sample was about 8 mg. The samples were first cooled from room temperature to -60 ℃ at a cooling rate of 10 ℃/min. Afterwards the heat flow of the samples was recorded at a heating rate of 10 ℃/min. The glass transition temperature ($T_g$) was determined as the temperature at the inflexion point of the heat flow curve. The $T_g$ values of the elastomers with HMDA/DADD ratios of 0.3/0.7 and 0.5/0.5 are 4 ℃ and 14 ℃ respectively.

**S8. Mechanical tests.** We characterized the mechanical properties of hybrid elastomers using Instron® 3342 with a 1000N load cell. Dumbbell shaped samples were cut from the sheets formed in the Teflon mold using a normalized cutter. The dumbbell shaped samples have a central part of 10mm in length, 4 mm in width and 0.8-1.2 mm in thickness. Uniaxial tensile measurements were conducted at room temperature in air using different strain rates, including 0.0014, 0.014, 0.14, 0.41 and 0.68 /sec. Engineering stress is defined as the tensile force per unit of initial cross-section area of the central part. Strain was recorded by a camera, which monitored the displacement of two markers in the central part of a dumbbell shaped sample. Each measurement was repeated at least three times.

Cyclic extension tests with incremental elongation were performed on the same tensile machine at a strain rate of 0.014/sec. The maximum strains of the cyclic extension tests included 0.3, 0.5, 0.8, 1.1 and 1.4. The strain was also recorded and measured using a camera. The efficiency of energy dissipation during each cyclic extension test is defined as the ratio of the integrated area in the hysteresis loop to that under the loading curve. The efficiency is remarkably high with the value of ~75%, as shown in **Figure S7**. Moreover, it is nearly independent of elongation.

**S9. Fracture toughness.** Fracture tests were conducted using the classical single edge notch test on the Instron® 3342 with a 1000N load cell. A notch of 1 mm in length was made in the middle of a rectangular specimen of about 1 mm in thickness and 5 mm in width. The specimen was fixed in the two clamps with a pre-set distance of 10 mm. Then, the specimen was subjected to uniaxial tension with a crosshead velocity of 3 mm/min, corresponding to a nominal strain rate of 0.005 s$^{-1}$. The strain at break ($\lambda_c$) of the notched specimen was recorded with the camera. The fracture energy ($G_c$) was calculated by a method developed by Greensmith:[1]

$$G_\mathrm{c} = \frac{6WC}{\sqrt{\lambda_\mathrm{c}}}$$

where c is the notch length. *W* is the strain energy calculated by integration of the stress-strain curve of an un-notched specimen until $\lambda_c$; the un-notched specimen underwent a tension process with the same strain rate as the notched sample.

**SI Figures**

Step 1: Michael addition

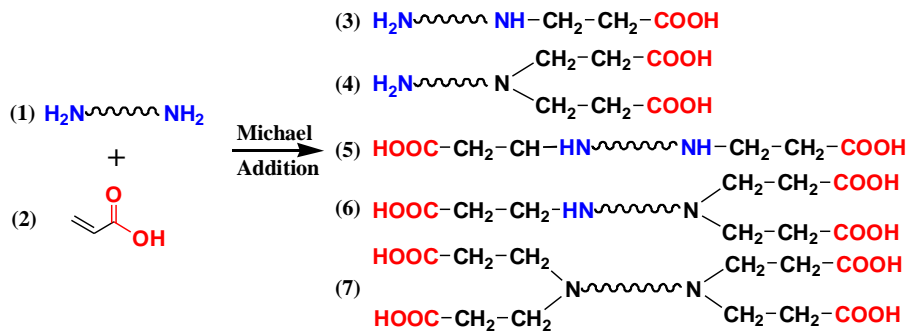

Step 2: Polycondensation

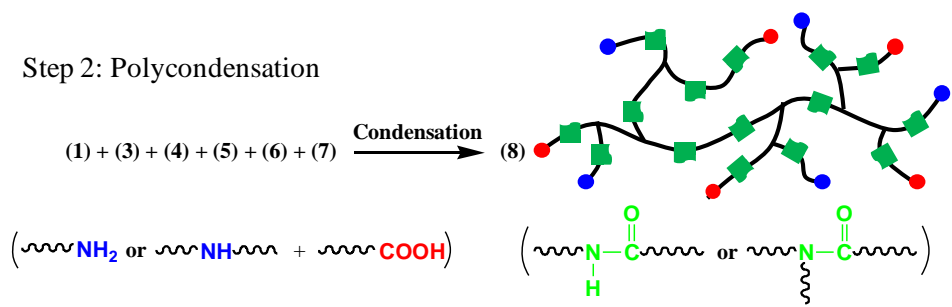

**Figure S1**. **Synthesis of randomly branched polymer.** Step 1: Reaction between diamine and acrylic acid produces five types of oligomers by Michael addition at 50°C. Step 2: Polycondensation reaction between amine and carboxyl groups 150 °C gives rise to a randomly branched polymer.

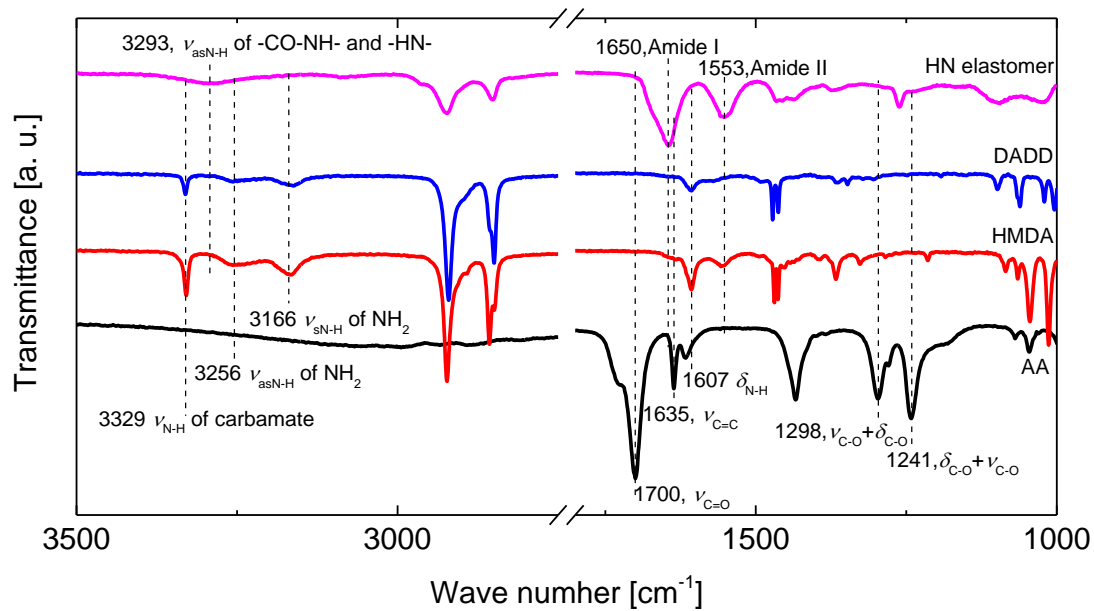

**Figure S2.** FTIR spectra of AA, HMDA, DADD and hybrid network.

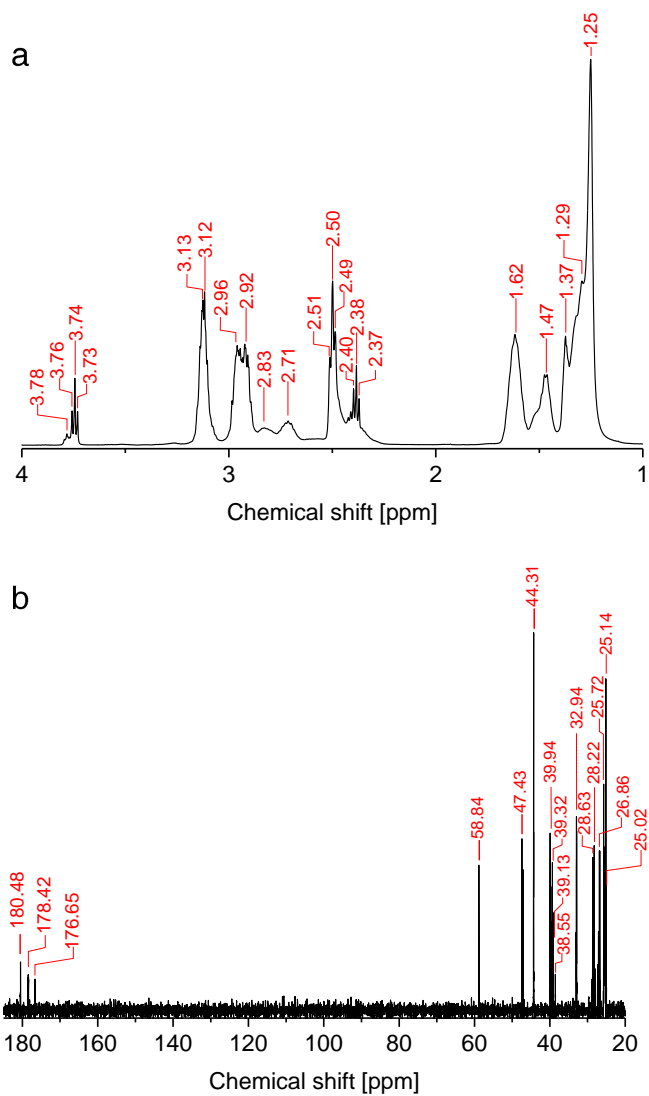

**Figure S3.** (a) $^1$H NMR and (b) $^{13}$C NMR spectra of the pre-polymer.

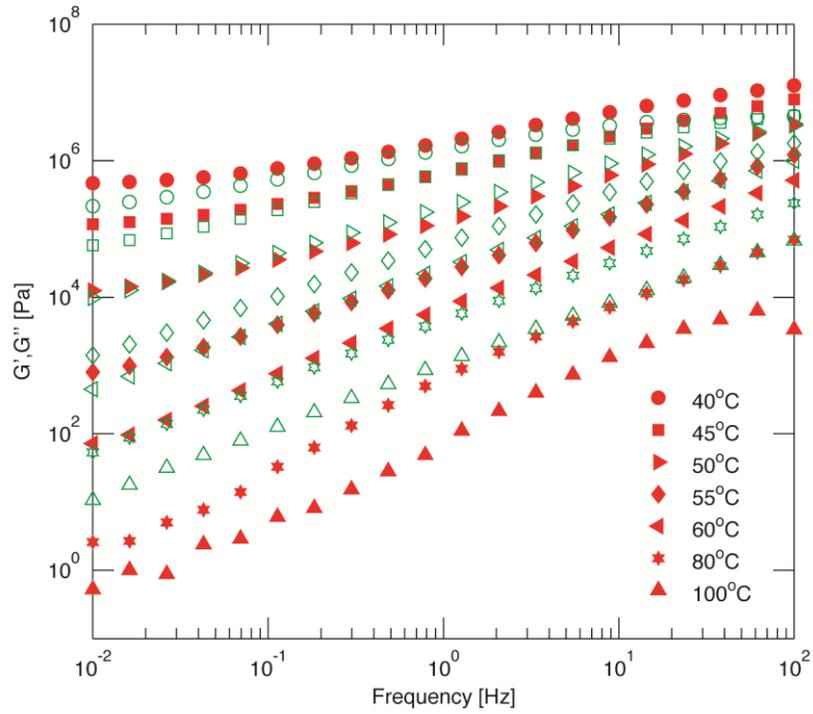

**Figure S4**. Dependence of storage modulus G' (filled symbols), and loss modulus G'' (empty symbols) on frequency for the melt of pre-polymer with a HMDA/DADD ratio of 0.7/0.3 under different temperatures.

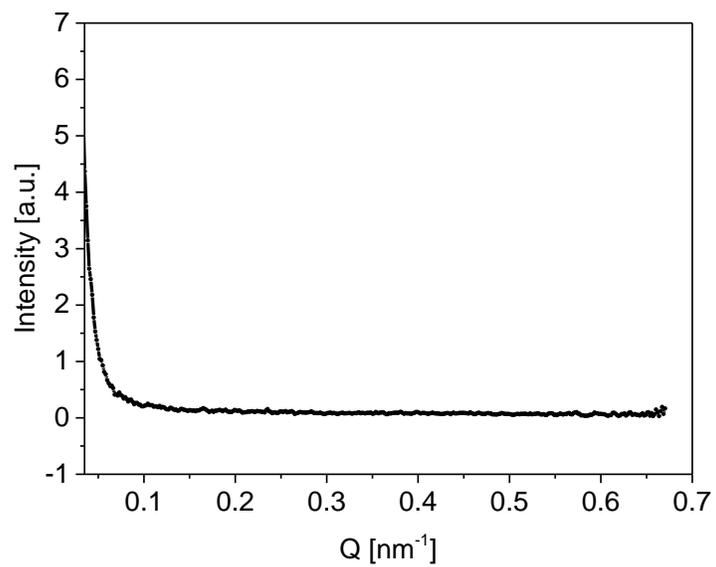

**Figure S5**. SAXS intensity as a function of scattering vector (Q) based on 360° azimuthally integration of 2D images.

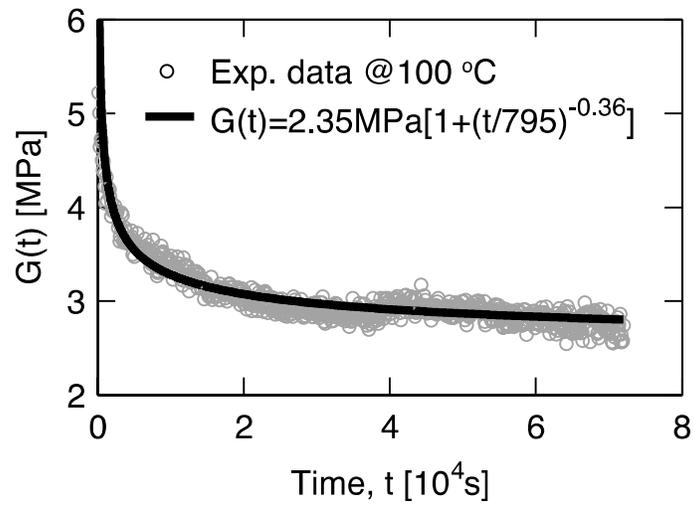

**Figure S6**. Stress relaxation of the hybrid network at 100°C. Dots are experimental data, and the solid line represents power law fit to the experimental data.

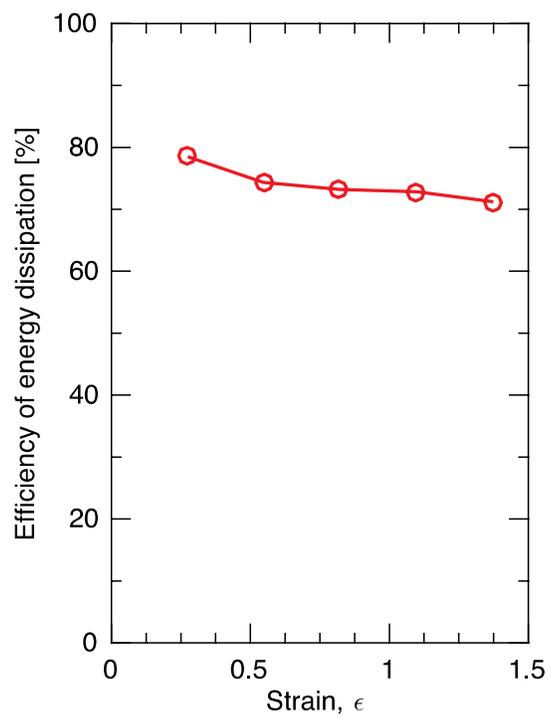

**Figure S7**. Dependence of energy dissipation efficiency on extents of strain.

**SI Movies**

**SI movie 1**. Uniaxial tensile test of a hybrid elastomer sample at strain rate of 0.014/sec. The movie is sped up by 5 times.